\documentclass[12pt]{article}
\usepackage{epsf}
\usepackage{graphicx}
\usepackage{amssymb}
\def\uu{\langle \bar u u \rangle}
\def\dd{\langle \bar d d \rangle}
\def\ss{\langle \bar s s \rangle}
\def\qq{\langle \bar q q \rangle}

\title{ {Magnetic Moments of Heavy $\Xi_{Q}$ Baryons in Light Cone QCD Sum Rules  }}
\author{
  T. M. Aliev \thanks{ e-mail: taliev@metu.edu.tr},
  K. Azizi\thanks {e-mail:e146342@metu.edu.tr},
  A. Ozpineci \thanks {e-mail:ozpineci@metu.edu.tr} \\
  \small Physics
Department, Middle East Technical University, 06531, Ankara, Turkey }
 \date{}
\begin{document}
\setlength{\baselineskip}{24pt} \maketitle
\setlength{\baselineskip}{7mm}
\begin{abstract}
The magnetic moments of heavy  $\Xi_{Q}$ baryons  containing a single charm or
bottom quark are calculated in the framework of light cone QCD sum
rules method. A comparison of our results with the predictions of other approches, such as relativistic and nonrelativistic  quark models, hyper central model, Chiral perturbation theory, soliton and skyrmion models is presented.
\end{abstract}
PACS: 11.55.Hx, 13.40.Em, 14.20.Lq, 14.20.Mr
\thispagestyle{empty}
\newpage
\setcounter{page}{1}
\section{Introduction}
During the last few years, exciting experimental results are obtained in
the baryon sector containing a single b-quark. The CDF Collaboration
 observed the states $\Sigma^{\pm}_{b}$ and $\Sigma^{\ast\pm}_{b}$
\cite{Aaltonen1}, while both DO \cite{Abazov} and CDF \cite
{Aaltonen2} Collaborations have  seen $\Xi_{b}$. Recently, BaBar
Collaboration reported the discovery of $\Omega^{\ast}_{c}$ with
mass splitting
$m_{\Omega^{\ast}_{c}}-m_{\Omega_{c}}=(70.8\pm1.0\pm1.1) MeV$
\cite{Aubert}.

The masses of the heavy baryons have been studied in the framework of
various phenomenological models \cite{Capstick}- \cite{Karliner3}
and also in the framework of QCD sum rules method \cite{Shifman}-
\cite{Liu}. Along  with their masses, another static parameter of the heavy
baryons is their magnetic moment. Study of the magnetic moments
can give valuable information about the  internal structures of hadrons.

The magnetic moments of heavy baryons have been studied in the framework
of different methods. In \cite{ Choudhury, Lic} the magnetic moments
of charmed baryons are calculated within naive quark model. In \cite{
Glozman, Julia}, magnetic moments of charmed and bottom baryons are
analyzed in quark model and in \cite{Schollamd} heavy baryon
magnetic moments are studied in bound state approach.
Magnetic moments of heavy baryons  are calculated in the relativistic three-quark
model  \cite{Faessler},  hyper central model \cite{Patel}, Chiral perturbation model \cite{Savage},
 soliton model  \cite{DOR}, skyrmion model \cite{Oh}  and nonrelativistic constituent quark model with light and strange $\bar qq$ pairs \cite{An}. In \cite{Zhu2} the magnetic
moments of $\Sigma_{c}$ and $\Lambda_{c}$ baryons are calculated in
QCD sum rules in external electromagnetic field. In \cite{Aliev1,
Aliev2}, the light cone QCD sum rules method is applied to study the
magnetic moments of the $\Lambda_{Q}$, $(Q=c,b)$ and
$\Sigma_{Q}\Lambda_{Q}$ transitions  (more about this
method can be found in \cite{Colangelo , Braun1, Balitsky, Braun2}  and
references therein ).

The aim of  the present work is the calculation of the magnetic moments of
the $\Xi_{b}$ baryons recently observed  by DO and CDF Collaborations
within the light cone QCD sum rules framework. The plan of the paper is as follows.
In section 2, using the general form of the the baryon current, the
light cone QCD sum rules for $\Xi_{b}$ and $\Xi_{c}$ baryons are
calculated. In section 3 we present our numerical calculations on
the  $\Xi_{b}$ and $\Xi_{c}$ baryons. In this section we also
present a comparison of our results with the predictions of other approaches.
\section{Light cone QCD sum rules for the $\Xi_{Q}$ magnetic moments  }
In order to calculate the magnetic moments of $\Xi_{Q}$ $(Q=b, c)$
in the framework of the light cone QCD sum rules, we need the expression for
the interpolating current of $\Xi_{Q}$. To construct it, we follow
\cite{Karliner2}, i.e. we assume that the strange and light quarks (sq) in
$\Xi_{Q}$ are in a relative spin zero state (scalar or pseudo scalar
diquarks). Therefore, the most general current without derivatives and  
with the quantum numbers of  $\Xi_{Q}$ can be constructed from the 
combination of aforementioned scalar or pseudoscalar diquarks in the 
following way
\begin{equation}\label{cur.xi}
\eta_{Q}=\varepsilon_{abc}[(q^{aT}Cs^{b})\gamma_{5}+\beta(q^{aT}C\gamma_{5}s^{b})]Q^{c},
\end{equation}
here a, b and c are color indices, C is the charge conjugation
operator, $Q=b$, or $c$,   $q=u$, or $d$ and $\beta$ is an arbitrary parameter.
Having the explicit expression for the interpolating current, our
next task is to construct light cone QCD sum rules for the magnetic
moments of $\Xi_{Q}$ baryons. It is constructed from the following
correlation function:
\begin{equation}\label{T}
\Pi(p,q)=i\int d^{4}xe^{ipx}<\gamma\mid
T\{\eta_{Q}(x)\bar{\eta}_{Q}(0)\mid \}0>.
\end{equation}

 The calculation of the phenomenological side at the hadronic level proceeds by inserting into  the correlation function a
  complete set of hadronic states with the quantum numbers of $\Xi_{Q}$. We get
\begin{equation}\label{corr.func1}
\Pi=\sum_{i}\frac{<0\mid \eta_{Q}\mid
\Xi_{Q_{i}}(p_{2})>}{p_{2}^{2}-m_{\Xi_{Q}}^{2}}<\Xi_{Q_{i}}(p_{2})\mid
\Xi_{Q_{i}}(p_{1})>_\gamma \frac{<\Xi_{Q_{i}}(p_{1})\mid
\bar\eta_{Q}\mid 0>}{p_{1}^{2}-m_{\Xi_{Q}}^{2}}.
\end{equation}
  Isolating  the ground state's contributions, Eq. (\ref{corr.func1}) can be written as
\begin{eqnarray}\label{corr.func2}
\Pi&=&\frac{<0\mid \eta_{Q}\mid
\Xi_{Q}(p_{2})>}{p_{2}^{2}-m_{\Xi_{Q}}^{2}}<\Xi_{Q}(p_{2})\mid
\Xi_{Q}(p_{1})>_\gamma\frac{<\Xi_{Q}(p_{1})\mid
\bar\eta_{Q}\mid 0>}{p_{1}^{2}-m_{\Xi_{Q}}^{2}}\nonumber\\
&+& \sum_{h_{i}}\frac{<0\mid \eta_{Qi}\mid
h_{i}(p_{2})>}{p_{2}^{2}-m_{h_{i}}^{2}}<h_{i}(p_{2})\mid
h_{i}(p_{1})>_\gamma\frac{<h_{i}(p_{1})\mid
\bar\eta_{Q}\mid 0>}{p_{1}^{2}-m_{h_{i}}^{2}},\nonumber\\
\end{eqnarray}
where $p_{1}=p+q$,  $p_{2}=p$ and q is the photon momentum. The second term in
Eq. (\ref{corr.func2}) describes the higher resonances and continuum
contributions. The coupling of the interpolating current with the
baryons $\Xi_{Q}$ is determined  as
\begin{equation}\label{lambda}
<0\mid \eta_{Q}\mid \Xi_{Q}(p)>=\lambda_{Q}u_{\Xi_{Q}}(p),
\end{equation}
where $u_{\Xi_{Q}}(p)$ is a spinor describing the  baryon $\Xi_{Q}$ with four momentum p and $\lambda_{Q}$ is the corresponding residue. 

    The last step for obtaining the expression for the physical part of the correlator function is to write down the matrix
    element $<\Xi_{Q}(p_{2})\mid
\Xi_{Q}(p_{1})>_{\gamma}$ in terms of the form factors. Using Lorentz covariance,
this matrix element can be written  as
\begin{eqnarray}\label{mat.el}
<\Xi_{Q}(p_{1})\mid \Xi_{Q}(p_{2})>_{\gamma}&=&
\varepsilon^{\mu}\overline{u}_{\Xi_{Q}}(p_{1})\left[f_{1}\gamma_{\mu}-i\frac{\sigma_{\mu\alpha}q_{\alpha}}{2m_{\Xi_{Q}}}f_{2}\right]u_{\Xi_{Q}}(p_{2})\nonumber\\&=&
\overline{u}_{\Xi_{Q}}(p_{1})\left[(f_{1}+f_{2})\gamma_{\mu}+\frac{(p_{1}+p_{2})_{{\mu}}}{2m_{\Xi_{Q}}}f_{2}\right]u_{\Xi_{Q}}(p_{2})\varepsilon^{\mu},\nonumber\\
\end{eqnarray}
where $f_{1}(q^{2})$ and $f_{2}(q^{2})$ are the form factors and
$\varepsilon^{\mu}$ is the photon polarization vector.

For calculation of the $\Xi_{Q}$ magnetic moments, the values of the
form factors only at $q^2=0$ are needed because the photon is  real
in our problem. Using Eqs. (\ref{corr.func2}-\ref{mat.el}) for
physical part of the correlator and summing over the spins of initial and final $\Xi_{Q}$ baryons, the correlation function becomes
\begin{eqnarray}\label{corr.func3}
\Pi&=&-\lambda_{Q}^{2}\varepsilon^{\mu}\frac{\not\!p_{2}+m_{\Xi_{Q}}}{p_{2}^{2}-m_{\Xi_{Q}}^{2}}
\left[(f_{1}+f_{2})\gamma_{\mu}+\frac{(p_{1}+p_{2})_{{\mu}}}{2m_{\Xi_{Q}}}f_{2}\right]\frac{\not\!p_{1}+m_{\Xi_{Q}}}{p_{1}^{2}-m_{\Xi_{Q}}^{2}}.\nonumber\\
\end{eqnarray}
From this expression, we see that there are various structures which
can be chosen for studying the magnetic moments of $\Xi_{Q}$. In the present work following \cite{ref}, we choose the structure
$\not\!p_{2}\not\!\varepsilon\not\!q$ that contains magnetic form
factor $f_{1}+f_{2}$ and at $q^2=0$ it gives the magnetic moment of
$\Xi_{Q}$ in units of $e\hbar/2m_{\Xi_{Q}}$. Choosing this structure
in the physical part of the correlator, for the magnetic moments of
$\Xi_{Q}$ we obtain
\begin{eqnarray}\label{corr.func4}
\Pi&=&-\lambda_{Q}^{2}\frac{1}{p_{1}^{2}-m_{\Xi_{Q}}^{2}}
\mu_{\Xi_{Q}}\frac{1}{p_{2}^{2}-m_{\Xi_{Q}}^{2}},\nonumber\\
\end{eqnarray}
where $\mu_{\Xi_{Q}}=(f_{1}+f_{2})\mid_{q^2=0}$ are the magnetic
moments of $\Xi_{Q}$  in units of $e\hbar/2m_{\Xi_{Q}}$.

In order to calculate the magnetic moments of $\Xi_{Q}$ baryons, the
expression of the theoretical part of the correlation function is
needed. After simple calculations for the theoretical part of the
correlation function in QCD we obtain
\begin{eqnarray}\label{mut.m}
\Pi&=&-i\epsilon_{abc}\epsilon_{a'b'c'}\int
d^{4}xe^{ipx}\langle\gamma(q)\mid\{\gamma_{5}S_{Q}^{cc'}
\gamma_{5}Tr(S_{q}^{ba'}S'^{ab'}_{s})\nonumber\\&+& \beta\gamma_{5}S_{Q}^{cc'}
Tr(S_{q}^{ba'}\gamma_{5}S'^{ab'}_{s})+\beta S_{Q}^{cc'}
\gamma_{5}Tr(\gamma_{5}S_{q}^{ba'}S'^{ab'}_{s})\nonumber\\&+& \beta^{2}S_{Q}^{cc'}
Tr(\gamma_{5}S^{ab'}_{s}\gamma_{5}S'^{ba'}_{q})\}\mid 0\rangle
\end{eqnarray}
where $S'_{i}=CS^{T}_{i}C$ and C and T are the charge conjugation and
transposition operators, respectively and  $S_{Q}$ and $S_{q(s)}$ are the
heavy and light(strange) quark propagators.

The correlation function from QCD part receives three different
contributions: a) perturbative contributions, b) nonperturbative
contributions, where photon is emitted from the freely propagating
quark (in other words at short distance) c) nonperturbative
contributions, when photon is radiated at long distances. To obtain the expression for the contribution from the emission of photon at short distances, the following procedure can be used: Each one of the quarks can emit the photon, and hence each term in Eq. (\ref{mut.m})  corresponds to three terms in which the propagator of the photon emitting quark is replaced by:
\begin{equation}\label{rep1}
S^{ab}_{\alpha \beta} \Rightarrow- \frac{1}{2} \left\{ \int d^4 y
F^{\mu \nu} y_\nu S^{free} (x-y) \gamma_\mu
S^{free}(y)\right\}^{ab}_{\alpha \beta} ,
\end{equation}
where the Fock-Schwinger gauge,
$x_\mu A^\mu(x)=0$ has been used. Note that the explicit expressions
of free light and heavy quark propagators in x representation are:
\begin{eqnarray}\label{free1}
S^{free}_{q} &=&\frac{i\not\!x}{2\pi^{2}x^{4}}-\frac{m_{q}}{4\pi^{2}x^{2}},\nonumber\\
S^{free}_{Q}
&=&\frac{m_{Q}^{2}}{4\pi^{2}}\frac{K_{1}(m_{Q}\sqrt{-x^2})}{\sqrt{-x^2}}-i
\frac{m_{Q}^{2}\not\!x}{4\pi^{2}x^2}K_{2}(m_{Q}\sqrt{-x^2}),\nonumber\\
\end{eqnarray}
where $K_{i}$ are Bessel functions, $m_{u, d}=0$ and $m_{s}\neq0$.
 The expression for the nonperturbative contributions to the
 correlation function can be obtained from Eq.
(\ref{mut.m}) by replacing one of the light quark propagator by
\begin{equation}\label{rep2}
\label{rep} S^{ab}_{\alpha \beta} \rightarrow - \frac{1}{4} \bar q^a
\Gamma_j q^b ( \Gamma_j )_{\alpha \beta}~,
\end{equation}
 where $\Gamma_j = \Big\{
1,~\gamma_5,~\gamma_\alpha,~i\gamma_5 \gamma_\alpha, ~\sigma_{\alpha
\beta} /\sqrt{2}\Big\}$ and sum over $\Gamma_j$ is implied, and the
other two propagators are the full propagators involving both
perturbative and nonperturbative contributions. In order to calculate
the correlation function from QCD side, we need the explicit
expressions of the heavy and light quark propagators in the presence
of external field.

The light cone expansion of the propagator in external field is
obtained  in \cite{Balitsky}. It receives contributions from various $\bar q
G q$, $\bar q G G q$, $\bar q q \bar q q$ nonlocal operators , where
$G$ is the gluon field strength tensor. In this work, we consider
operators with only one gluon field and contributions coming from
 three particle nonlocal operators and neglect terms with two
gluons $\bar q GG q$, and four quarks $\bar q q \bar q q$
 \cite{Braun2}. In this approximation the  expressions for the heavy  and
light quark propagators are:
\begin{eqnarray}\label{heavylight}
i S_Q (x)& =& i S_Q^{free} (x) - i g_s \int \frac{d^4 k}{(2\pi)^4}
e^{-ikx} \int_0^1 dv \Bigg[\frac{\not\!k + m_Q}{( m_Q^2-k^2)^2}
G^{\mu\nu}(vx)
\sigma_{\mu\nu} \nonumber \\
&+& \frac{1}{m_Q^2-k^2} v x_\mu G^{\mu\nu} \gamma_\nu \Bigg],
\nonumber \\
S_q(x) &=&  S_q^{free} (x) - \frac{m_q}{4 \pi^2 x^2} - \frac{\langle
\bar q q \rangle}{12} \left(1 - i \frac{m_q}{4} \not\!x \right) -
\frac{x^2}{192} m_0^2 \langle \bar q q \rangle \left( 1 - i
\frac{m_q}{6}\not\!x \right) \nonumber \\ &&
 - i g_s \int_0^1 du \left[\frac{\not\!x}{16 \pi^2 x^2} G_{\mu \nu} (ux) \sigma_{\mu \nu} - u x^\mu
G_{\mu \nu} (ux) \gamma^\nu \frac{i}{4 \pi^2 x^2} \right. \nonumber
\\ && \left. - i \frac{m_q}{32 \pi^2} G_{\mu \nu} \sigma^{\mu \nu}
\left( \ln \left( \frac{-x^2 \Lambda^2}{4} \right) + 2 \gamma_E
\right) \right],
 \end{eqnarray}
 where  $\Lambda$ is the energy cut off separating
 perturbative and nonperturbative domains.

 In order to calculate the theoretical part, from Eqs.
 (\ref{mut.m})-(\ref{heavylight}) it follows that the matrix
 elements of nonlocal operators $\bar q \Gamma_{i}q$ between the photon
 and vacuum states are needed, i.e.$ \langle\gamma(q)\mid\bar q(x_{1})
 \Gamma_{i}q(x_{2})\mid0\rangle$. These matrix elements can be
 expanded near the light cone $x^2=0$  in terms of
 the photon distribution amplitudes \cite{Ball}.
 \begin{eqnarray}
&&\langle \gamma(q) \vert  \bar q(x) \sigma_{\mu \nu} q(0) \vert  0
\rangle  = -i e_q \bar q q (\varepsilon_\mu q_\nu - \varepsilon_\nu
q_\mu) \int_0^1 du e^{i \bar u qx} \left(\chi \varphi_\gamma(u) +
\frac{x^2}{16} \mathbb{A}  (u) \right) \nonumber \\ &&
-\frac{i}{2(qx)}  e_q \qq \left[x_\nu \left(\varepsilon_\mu - q_\mu
\frac{\varepsilon x}{qx}\right) - x_\mu \left(\varepsilon_\nu -
q_\nu \frac{\varepsilon x}{q x}\right) \right] \int_0^1 du e^{i \bar
u q x} h_\gamma(u)
\nonumber \\
&&\langle \gamma(q) \vert  \bar q(x) \gamma_\mu q(0) \vert 0 \rangle
= e_q f_{3 \gamma} \left(\varepsilon_\mu - q_\mu \frac{\varepsilon
x}{q x} \right) \int_0^1 du e^{i \bar u q x} \psi^v(u)
\nonumber \\
&&\langle \gamma(q) \vert \bar q(x) \gamma_\mu \gamma_5 q(0) \vert 0
\rangle  = - \frac{1}{4} e_q f_{3 \gamma} \epsilon_{\mu \nu \alpha
\beta } \varepsilon^\nu q^\alpha x^\beta \int_0^1 du e^{i \bar u q
x} \psi^a(u)
\nonumber \\
&&\langle \gamma(q) | \bar q(x) g_s G_{\mu \nu} (v x) q(0) \vert 0
\rangle = -i e_q \qq \left(\varepsilon_\mu q_\nu - \varepsilon_\nu
q_\mu \right) \int {\cal D}\alpha_i e^{i (\alpha_{\bar q} + v
\alpha_g) q x} {\cal S}(\alpha_i)
\nonumber \\
&&\langle \gamma(q) | \bar q(x) g_s \tilde G_{\mu \nu} i \gamma_5 (v
x) q(0) \vert 0 \rangle = -i e_q \qq \left(\varepsilon_\mu q_\nu -
\varepsilon_\nu q_\mu \right) \int {\cal D}\alpha_i e^{i
(\alpha_{\bar q} + v \alpha_g) q x} \tilde {\cal S}(\alpha_i)
\nonumber \\
&&\langle \gamma(q) \vert \bar q(x) g_s \tilde G_{\mu \nu}(v x)
\gamma_\alpha \gamma_5 q(0) \vert 0 \rangle = e_q f_{3 \gamma}
q_\alpha (\varepsilon_\mu q_\nu - \varepsilon_\nu q_\mu) \int {\cal
D}\alpha_i e^{i (\alpha_{\bar q} + v \alpha_g) q x} {\cal
A}(\alpha_i)
\nonumber \\
&&\langle \gamma(q) \vert \bar q(x) g_s G_{\mu \nu}(v x) i
\gamma_\alpha q(0) \vert 0 \rangle = e_q f_{3 \gamma} q_\alpha
(\varepsilon_\mu q_\nu - \varepsilon_\nu q_\mu) \int {\cal
D}\alpha_i e^{i (\alpha_{\bar q} + v \alpha_g) q x} {\cal
V}(\alpha_i) \nonumber \\ && \langle \gamma(q) \vert \bar q(x)
\sigma_{\alpha \beta} g_s G_{\mu \nu}(v x) q(0) \vert 0 \rangle  =
e_q \qq \left\{
        \left[\left(\varepsilon_\mu - q_\mu \frac{\varepsilon x}{q x}\right)\left(g_{\alpha \nu} -
        \frac{1}{qx} (q_\alpha x_\nu + q_\nu x_\alpha)\right) \right. \right. q_\beta
\nonumber \\ && -
         \left(\varepsilon_\mu - q_\mu \frac{\varepsilon x}{q x}\right)\left(g_{\beta \nu} -
        \frac{1}{qx} (q_\beta x_\nu + q_\nu x_\beta)\right) q_\alpha
\nonumber \\ && -
         \left(\varepsilon_\nu - q_\nu \frac{\varepsilon x}{q x}\right)\left(g_{\alpha \mu} -
        \frac{1}{qx} (q_\alpha x_\mu + q_\mu x_\alpha)\right) q_\beta
\nonumber \\ &&+
         \left. \left(\varepsilon_\nu - q_\nu \frac{\varepsilon x}{q.x}\right)\left( g_{\beta \mu} -
        \frac{1}{qx} (q_\beta x_\mu + q_\mu x_\beta)\right) q_\alpha \right]
   \int {\cal D}\alpha_i e^{i (\alpha_{\bar q} + v \alpha_g) qx} {\cal T}_1(\alpha_i)
\nonumber \\ &&+
        \left[\left(\varepsilon_\alpha - q_\alpha \frac{\varepsilon x}{qx}\right)
        \left(g_{\mu \beta} - \frac{1}{qx}(q_\mu x_\beta + q_\beta x_\mu)\right) \right. q_\nu
\nonumber \\ &&-
         \left(\varepsilon_\alpha - q_\alpha \frac{\varepsilon x}{qx}\right)
        \left(g_{\nu \beta} - \frac{1}{qx}(q_\nu x_\beta + q_\beta x_\nu)\right)  q_\mu
\nonumber \\ && -
         \left(\varepsilon_\beta - q_\beta \frac{\varepsilon x}{qx}\right)
        \left(g_{\mu \alpha} - \frac{1}{qx}(q_\mu x_\alpha + q_\alpha x_\mu)\right) q_\nu
\nonumber \\ &&+
         \left. \left(\varepsilon_\beta - q_\beta \frac{\varepsilon x}{qx}\right)
        \left(g_{\nu \alpha} - \frac{1}{qx}(q_\nu x_\alpha + q_\alpha x_\nu) \right) q_\mu
        \right]
    \int {\cal D} \alpha_i e^{i (\alpha_{\bar q} + v \alpha_g) qx} {\cal T}_2(\alpha_i)
\nonumber \\ &&+
        \frac{1}{qx} (q_\mu x_\nu - q_\nu x_\mu)
        (\varepsilon_\alpha q_\beta - \varepsilon_\beta q_\alpha)
    \int {\cal D} \alpha_i e^{i (\alpha_{\bar q} + v \alpha_g) qx} {\cal T}_3(\alpha_i)
\nonumber \\ &&+
        \left. \frac{1}{qx} (q_\alpha x_\beta - q_\beta x_\alpha)
        (\varepsilon_\mu q_\nu - \varepsilon_\nu q_\mu)
    \int {\cal D} \alpha_i e^{i (\alpha_{\bar q} + v \alpha_g) qx} {\cal T}_4(\alpha_i)
                        \right\}
\end{eqnarray}
where $\chi$ is the magnetic susceptibility of the quarks,
$\varphi_\gamma(u)$ is the leading twist 2, $\psi^v(u)$,
$\psi^a(u)$, ${\cal A}$ and ${\cal V}$ are the twist 3 and
$h_\gamma(u)$, $\mathbb{A}$, ${\cal T}_i$ ($i=1,~2,~3,~4$) are the
twist 4 photon distribution amplitudes (DA's), respectively. The explicit expressions of DA's are presented in numerical analysis section.

The theoretical part of the correlation function can be obtained in
terms of QCD parameters by substituting photon DA's and expressions
for heavy and light quarks propagators in to Eq. (\ref{mut.m}). Sum
rules for the $\Xi_{Q}$ magnetic moments are obtained by equating
two representations of the correlation function. The higher states
and continuum contributions are modeled using the hadron-quark
duality. Applying double Borel transformations on the variables
$p_{1}^{2}=(p+q)^{2}$ and $p_{2}^{2}=p^{2}$ on  both sides of the
correlator, for the $\Xi_{Q}$ baryon magnetic moments we get:
\begin{eqnarray}\label{buneya}
&-&\lambda_{Q}^{2}(\beta)\mu_{\Xi_{Q}}e^{-m_{\Xi_{Q}}^{2}/M_{B}^{2}}=\int_{m_{Q}^{2}}^{s_{0}}e^{\frac{-s}{M_{B}^{2}}}\rho(s)ds\nonumber\\&+&\frac{(\beta^2-1)e^{\frac{-m_{Q}^{2}}{M_{B}^{2}}}}{288\pi^{2}}
\left\{\vphantom{\int_0^{x_2}}\gamma_{E}(6e_{Q}+e_{s})m_{s}m_{0}^{2}<\bar
qq>\vphantom{\int_0^{x_2}}\right\}\nonumber\\&-&\frac{(\beta^2-1)e^{\frac{-m_{Q}^{2}}{M_{B}^{2}}}m_{Q}^2}
{72M_{B}^{4}}\left\{(e_{s}+e_{q})m_{0}^{2}<\bar
ss><\bar
qq>\eta_{1}\vphantom{\int_0^{x_2}}\right\}\nonumber\\
&+&\frac{e^{\frac{-m_{Q}^{2}}{M_{B}^{2}}}m_{Q}^{2}}{432M_{B}^{4}}\left\{\vphantom{\int_0^{x_2}}(\beta^2-1)m_{0}^{2}<\bar
ss><\bar qq>\left[\vphantom{\int_0^{x_2}}36e_{Q}+(e_{s}+e_{q})\mathbb{A}(u_{0})\vphantom{\int_0^{x_2}}\right]\right.\nonumber\\
&+&\left.(\beta^2+1)f_{3\gamma}e_{q}m_{s}m_{0}^{2}<\bar
ss>(\eta_{2}+\psi^{a}(u_{0}))\vphantom{\int_0^{x_2}}\right\}\nonumber\\
&+&\frac{e^{\frac{-m_{Q}^{2}}{M_{B}^{2}}}}{72}\left\{\vphantom{\int_0^{x_2}}(1-\beta^2)<\bar
ss><\bar qq>\left[\vphantom{\int_0^{x_2}}12e_{Q}-(e_{s}+e_{q})[\eta_{1}-m_{0}^{2}\chi_{i}\varphi_{\gamma}(u_{0})]\vphantom{\int_0^{x_2}}\right]\right.\nonumber\\
&-&3e_{q}(\beta^2+1)f_{3\gamma}m_{s}<\bar
ss>\eta_{2}\left.\vphantom{\int_0^{x_2}}\right\}\nonumber\\
&-&\frac{(\beta^2-1)e^{\frac{-m_{Q}^{2}}{M_{B}^{2}}}M_{B}^{2}m_{s}}{96\pi^{2}m_{Q}^{2}}
\left[\vphantom{\int_0^{x_2}}(6e_{Q}+e_{s})\gamma_{E}m_{0}^{2}<\bar qq>\vphantom{\int_0^{x_2}}\right]\nonumber\\
&-&
\frac{e^{\frac{-m_{Q}^{2}}{M_{B}^{2}}}m_{s}}{288\pi^2}\left(\vphantom{\int_0^{x_2}}3(\beta^2-1)e_{Q}m_{0}^{2}<\bar
qq>\left\{\vphantom{\int_0^{x_2}}-3+2\gamma_{E}+2ln[\frac{\Lambda^{2}}{m_{Q}^{2}}]\vphantom{\int_0^{x_2}}\right\}\right.\nonumber\\
&+&e_{q}m_{0}^{2}<\bar
ss>(1+\beta^2)+e_{s}m_{0}^{2}<\bar
qq>\left[\vphantom{\int_0^{x_2}}(1-\beta^2)(\gamma_{E}+ln[\frac{\Lambda^{2}}{m_{Q}^{2}}])\vphantom{\int_0^{x_2}}\right]
\left.\vphantom{\int_0^{x_2}}\right)\nonumber\\
&+&\frac{9e_{Q}m_{Q}^{2}m_{s}}{144\pi^2}\left\{\vphantom{\int_0^{x_2}}
<\bar ss>(1+\beta^2)+2<\bar qq>(1-\beta^2)\vphantom{\int_0^{x_2}}\right\}
,\nonumber\\
\end{eqnarray}
where $M_{B}^{2}=\frac{M_{1}^{2}M_{2}^{2}}{M_{1}^{2}+M_{2}^{2}}$ and
$u_{0}=\frac{M_{1}^{2}}{M_{1}^{2}+M_{2}^{2}}$.  Since the masses of
the initial and final baryons are the same, we will set
$M_{1}^{2}=M_{2}^{2}$ and $u_{0}=1/2$. The functions appearing in
Eq. (\ref{buneya}) are defined as:
\begin{eqnarray}\label{functions1}
\eta_{1} &=& \int {\cal D}\alpha_i \int_0^1 dv {\cal S}(\alpha_i)
\delta(\alpha_{ q} + v \alpha_g -  u_0),
\nonumber \\
\eta_{2} &=& \int {\cal D}\alpha_i \int_0^1 dv {\cal V}(\alpha_i)
\delta'(\alpha_{ q} + v \alpha_g -  u_0),
\nonumber \\
\rho(s)&=&\frac{(\beta^2-1)}
{144\pi^{2}M_{B}^{2}}\left\{\vphantom{\int_0^{x_2}}m_{0}^{2}(6e_{Q}+e_{s})m_{s}<\bar qq>\ln (\frac{-{m_{Q}}^2 + s}{{\Lambda}^2})\vphantom{\int_0^{x_2}}\right\}\nonumber\\
&+&\frac{3(1+\beta^2)e_{Q}m_{Q}^{4}}{64\pi^{4}}
\left\{\vphantom{\int_0^{x_2}}\frac{13}{2}+\psi_{10}-\frac{1}{6}\psi_{20}-\frac{1}{6}\psi_{30}
+[\psi_{10}+\frac{3}{2}]ln(\frac{s}{m_{Q}^{2}})+\frac{1}{6}\psi_{41}
\vphantom{\int_0^{x_2}}\right\}
\nonumber\\
&+&\frac{(1-\beta^2)m_{s}
<\bar qq>\psi_{10}}{48\pi^{2}m_{Q}^{2}}\left\{\vphantom{\int_0^{x_2}}2e_{q}m_{Q}^{2}\eta_{1}-(e_{s}+e_{q})
m_{0}^{2}\left[\vphantom{\int_0^{x_2}}8+ln(\frac{s-m_{Q}^{2}}{{\Lambda}^2})\vphantom{\int_0^{x_2}}\right]
\vphantom{\int_0^{x_2}}\right\}\nonumber\\
&-&\frac{m_{s}}{288\pi^{2}m_{Q}^{2}}\left\{\vphantom{\int_0^{x_2}}(\beta^2-1)m_{0}^{2}(e_{s}+6e_{Q})<\bar
qq>\right.\left[\vphantom{\int_0^{x_2}}3ln(\frac{s-m_{Q}^{2}}{{\Lambda}^2})-(4\gamma_{E}
+ln[\frac{\Lambda^{2}}{m_{Q}^{2}}])
\vphantom{\int_0^{x_2}}\right]\nonumber\\
&+&6e_{Q}\left[\vphantom{\int_0^{x_2}}3m_{Q}^{2}\{(1+\beta^2)<\bar
ss>+2(1-\beta^2)<\bar qq>\}\psi_{10}
\vphantom{\int_0^{x_2}}\right]\left.
\vphantom{\int_0^{x_2}}\right\}\nonumber\\
&+&\frac{m_{Q}^{2}}{576\pi^2}\left(\vphantom{\int_0^{x_2}}(e_{s}+12e_{Q})
\left\{\vphantom{\int_0^{x_2}}\right.\right.\frac{(\beta^2-1)}{m_{Q}^{4}}m_{0}^{2}m_{s}<\bar
qq>\left[\vphantom{\int_0^{x_2}}-(1+\gamma_{E})(\psi_{22}+2\psi_{12})\right.\nonumber\\
&-&\left.\left.\psi_{02}-\psi_{32}-\psi_{22}-\frac{\gamma_{E}}{2}\psi_{20}+3(2\psi_{32}+3\psi_{22}+\psi_{02})ln(\frac{s-m_{Q}^{2}}{{\Lambda}^2})-ln(\frac{\Lambda^2}{m_{Q}^2})
\vphantom{\int_0^{x_2}}\right]\vphantom{\int_0^{x_2}}\right\}\nonumber\\
&+&12e_{q}\left\{\vphantom{\int_0^{x_2}}\frac{2}{m_{Q}^{2}}(\beta^2-1)m_{s}<\bar
qq>\eta_{1}\psi_{21}\right.\nonumber\\
&+&\left.\left.(1+\beta^2)f_{3\gamma}\eta_{2}\left[\vphantom{\int_0^{x_2}}\psi_{21}-\psi_{10}+\frac{1}{2}\psi_{20}+\frac{1}{2}\psi_{00}
+ln(\frac{m_{Q}^{2}}{s})\vphantom{\int_0^{x_2}}\right]\vphantom{\int_0^{x_2}}\right\}
\vphantom{\int_0^{x_2}}\right),\nonumber \\
\end{eqnarray}
where,
\begin{eqnarray}\label{functions1}
\int {\cal D} \alpha_i& = &\int_0^1 d \alpha_{\bar q} \int_0^1 d
\alpha_q \int_0^1 d \alpha_g \delta(1-\alpha_{\bar
q}-\alpha_q-\alpha_g),
\end{eqnarray}
and functions $\psi_{nm}$ are defined as
\begin{eqnarray}\label{functions2}
\psi_{nm}&=&\frac{{( {s-m_{Q}}^2 )
}^n}{s^m{(m_{Q}^{2})}^{n-m}},\nonumber \\
\end{eqnarray}
Note that the contributions of the terms $\sim <G^2>$ are also calculated, but their numerical values are very small and therefore for customary in Eq. (\ref{buneya}) these terms are omitted.
From Eq. (\ref{buneya}) it follows that for the determination of the
$\Xi_{Q}$ baryon magnetic moments, we need to know the residue
$\lambda_{Q}$. The residue  can be obtained from the two-point sum
rules and  is calculated in \cite{Duraes}. For the current given in eq.
(\ref{cur.xi}) it takes the following form:
\begin{eqnarray}\label{resediue}
\lambda^2_{Q}(\beta)&=&e^{{m_{\Xi_{Q}}}^2/M_{B}^2}\left(
\vphantom{\int_0^{x_2}}\int_{m_Q^2}^{s_0}ds~ e^{-s/M_{B}^2}\left\{
\vphantom{\int_0^{x_2}}
 {(1+\beta^2)m_Q^4\over
2^{9}\pi^4}\left[(1-x^2)\left({1 \over x^2}-{8\over
x}+1\right)-12\ln{x}\right] \right.\right.\nonumber\\&+&
{m_s\over2^4\pi^2}(1-x^2)\left[(1-\beta^2)\qq+{(1+b^2)\ss \over2}\right]
\nonumber\\&+&\left. {(1+\beta^2)<g^{2}G^{2}>\over 2^{10}3
\pi^4}(1-x)(1+5x)\right\}
\nonumber\\
&+&{m_s\over2^5\pi^2}\left\{\frac{(1+\beta^2)m_{0}^{2}<\bar ss>}{6}
e^{-m_Q^2/M_{B}^2}-(1-\beta^2)m_{0}^{2}<\bar
qq>\left[e^{-m_Q^2/M_{B}^2}\right.\right.\nonumber\\
&+&\left.\left.\left.\int_0^1d\alpha(1-\alpha)
e^{-m_Q^2\over(1-\alpha)M_{B}^2}\right]\right\}
-\frac{\qq\ss}{6}(1-\beta^2)e^{-m_Q^2/M_{B}^2}\right),
\nonumber \\
\end{eqnarray}
where, $ x = m_{Q}^{2} /s $.
\section{Numerical analysis}
The present section is devoted to the numerical analysis of the  magnetic
moments of $\Xi_{Q}$ baryons. The values of the input parameters,
appearing in the sum rules expression for magnetic moments are:
$\uu(1~GeV) = \dd(1~GeV)= -(0.243)^3~GeV^3$, $\ss(1~GeV) = 0.8
\uu(1~GeV)$, $m_0^2(1~GeV) = 0.8~GeV^2$ \cite{Belyaev}, , $\Lambda =
300~MeV$ and $f_{3 \gamma} = - 0.0039~GeV^2$ \cite{Ball}. The value
of the magnetic susceptibility  $\chi(1~GeV)=-3.15\pm0.3~GeV^{-2}$ was
obtained by a combination of the local duality approach and QCD sum
rules  \cite{Ball}. Recently, from the analysis of radiative heavy meson
decay,  $\chi(1~GeV)=-(2.85\pm0.5)~GeV^{-2}$ was obtained
\cite{Rohrwild}, which is in good agreement with the instanton liquid model prediction \cite{Dorokov}, but slightly below  the QCD sum rules prediction \cite{Ball}. Note that firstly the magnetic susceptibility in the framework of QCD sum rules is calculated in \cite{Kogan} and it is obtained that $\chi(1~GeV)=-4.4~GeV^{-2}$. In the numerical analysis, we have used all three value of  $\chi$ existing  in the litrature and obtained that the values of the magnetic moments of $\Xi_{Q}$ baryons are practically insensitive to the value of $\chi$. The photon DA's entering  the sum rules for the magnetic moments of $\Xi_{Q}$ are calculated in \cite{Ball} and their expressions  are:
\begin{eqnarray}
\varphi_\gamma(u) &=& 6 u \bar u \left( 1 + \varphi_2(\mu)
C_2^{\frac{3}{2}}(u - \bar u) \right),
\nonumber \\
\psi^v(u) &=& 3 \left(3 (2 u - 1)^2 -1 \right)+\frac{3}{64} \left(15
w^V_\gamma - 5 w^A_\gamma\right)
                        \left(3 - 30 (2 u - 1)^2 + 35 (2 u -1)^4
                        \right),
\nonumber \\
\psi^a(u) &=& \left(1- (2 u -1)^2\right)\left(5 (2 u -1)^2 -1\right)
\frac{5}{2}
    \left(1 + \frac{9}{16} w^V_\gamma - \frac{3}{16} w^A_\gamma
    \right),
\nonumber \\
{\cal A}(\alpha_i) &=& 360 \alpha_q \alpha_{\bar q} \alpha_g^2
        \left(1 + w^A_\gamma \frac{1}{2} (7 \alpha_g - 3)\right),
\nonumber \\
{\cal V}(\alpha_i) &=& 540 w^V_\gamma (\alpha_q - \alpha_{\bar q})
\alpha_q \alpha_{\bar q}
                \alpha_g^2,
\nonumber \\
h_\gamma(u) &=& - 10 \left(1 + 2 \kappa^+\right) C_2^{\frac{1}{2}}(u
- \bar u),
\nonumber \\
\mathbb{A}(u) &=& 40 u^2 \bar u^2 \left(3 \kappa - \kappa^+
+1\right) \nonumber \\ && +
        8 (\zeta_2^+ - 3 \zeta_2) \left[u \bar u (2 + 13 u \bar u) \right.
\nonumber \\ && + \left.
                2 u^3 (10 -15 u + 6 u^2) \ln(u) + 2 \bar u^3 (10 - 15 \bar u + 6 \bar u^2)
        \ln(\bar u) \right],
\nonumber \\
{\cal T}_1(\alpha_i) &=& -120 (3 \zeta_2 + \zeta_2^+)(\alpha_{\bar
q} - \alpha_q)
        \alpha_{\bar q} \alpha_q \alpha_g,
\nonumber \\
{\cal T}_2(\alpha_i) &=& 30 \alpha_g^2 (\alpha_{\bar q} - \alpha_q)
    \left((\kappa - \kappa^+) + (\zeta_1 - \zeta_1^+)(1 - 2\alpha_g) +
    \zeta_2 (3 - 4 \alpha_g)\right),
\nonumber \\
{\cal T}_3(\alpha_i) &=& - 120 (3 \zeta_2 - \zeta_2^+)(\alpha_{\bar
q} -\alpha_q)
        \alpha_{\bar q} \alpha_q \alpha_g,
\nonumber \\
{\cal T}_4(\alpha_i) &=& 30 \alpha_g^2 (\alpha_{\bar q} - \alpha_q)
    \left((\kappa + \kappa^+) + (\zeta_1 + \zeta_1^+)(1 - 2\alpha_g) +
    \zeta_2 (3 - 4 \alpha_g)\right),\nonumber \\
{\cal S}(\alpha_i) &=& 30\alpha_g^2\{(\kappa +
\kappa^+)(1-\alpha_g)+(\zeta_1 + \zeta_1^+)(1 - \alpha_g)(1 -
2\alpha_g)\nonumber \\&+&\zeta_2
[3 (\alpha_{\bar q} - \alpha_q)^2-\alpha_g(1 - \alpha_g)]\},\nonumber \\
\tilde {\cal S}(\alpha_i) &=&-30\alpha_g^2\{(\kappa -
\kappa^+)(1-\alpha_g)+(\zeta_1 - \zeta_1^+)(1 - \alpha_g)(1 -
2\alpha_g)\nonumber \\&+&\zeta_2 [3 (\alpha_{\bar q} -
\alpha_q)^2-\alpha_g(1 - \alpha_g)]\}.
\end{eqnarray}
The constants appearing in the wave functions are given as
\cite{Ball} $\varphi_2(1~GeV) = 0$, $w^V_\gamma = 3.8 \pm 1.8$,
$w^A_\gamma = -2.1 \pm 1.0$, $\kappa = 0.2$, $\kappa^+ = 0$,
$\zeta_1 = 0.4$, $\zeta_2 = 0.3$, $\zeta_1^+ = 0$ and $\zeta_2^+ =
0$.

From the explicit expressions of the magnetic moments of $\Xi_{Q}$
baryons, it follows that it contains three auxiliary parameters:
Borel mass squared $M_{B}^{2}$, continuum threshold $s_{0}$ and $\beta$
which enters  the expression of the interpolating current for
$\Xi_{Q}$. The physical quantity, magnetic moment $\mu_{\Xi_{Q}}$,
should be independent of these auxiliary parameters. In other words we
should find the "working regions" of these auxiliary parameters,
where the magnetic moments are independent of them.

The value of the continuum threshold  is fixed from the analysis of the two-
point sum rules, where the mass and residue $\lambda_{\Xi_{Q}}$ of
the  $\Xi_{Q}$ baryons are determined  \cite{Duraes}, which leads to
the value $s_{0}=6.5^2~GeV^2$ for $\Xi_{b}$ and  $s_{0}=3.0^2~GeV^2$ for
$\Xi_{c}$.  If we choose the value $s_{0}=6.4^2~GeV^2$ for $\Xi_{b}$
and  $s_{0}=8~GeV^2$ for $\Xi_{c}$. the results remain practically
unchanged. Next, we try to find the working region of $M_{B}^{2}$
where $\mu_{\Xi_{Q}}$ are independent of it at fixed value of $\beta$
and the above mentioned  values of $s_{0}$. The upper bound of $M_{B}^{2}$
is obtained requiring that the continuum contribution should be less than
the contribution of the first resonance. The lower bound
of $M_{B}^{2}$ is determined by requiring that the highest power of
$1/M_{B}^{2}$ be less than   $30^0/_{0}$    of  the highest
power of $M_{B}^{2}$. These two conditions are both satisfied in the
region $15~GeV^2\leq M_{B}^{2}\leq20~GeV^2 $ and $5~GeV^2\leq
M_{B}^{2}\leq8~GeV^2 $ for $\Xi_{b}$ and $\Xi_{c}$, respectively.

 In Figs. 1 and 2, we depict the dependence of $\mu_{\Xi_{b}^{0}}$
 and $\mu_{\Xi_{b}^{-}}$ on $M_{B}^{2}$ at fixed value of $\beta$ and
 $s_{0}=6.5^2~GeV^2$. In Figs. 3 and 4, we present the dependence of $\mu_{\Xi_{c}^{0}}$
 and $\mu_{\Xi_{c}^{+}}$ on $M_{B}^{2}$ at fixed value of $\beta$ and
 $s_{0}=3.0^2~GeV^2$.  From these figures, we see that the values of the magnetic moments of $\Xi_{b}$ and $\Xi_{c}$ exhibit good stability when $M_{B}^{2}$ varies in the region $15~GeV^2\leq M_{B}^{2}\leq20~GeV^2 $ and $5~GeV^2\leq
M_{B}^{2}\leq8~GeV^2 $, respectively. The last step of our analysis is the determination of
 the working region for the auxiliary parameter $\beta$. For this aim,
 in Figs. 5, 6, 7, and 8 we present the dependence of the magnetic
 moments of $\Xi_{Q}$ baryons on $\cos\theta$ where $\tan\theta=\beta$, using the values of
 $M_{B}^{2}$ from the "working region" which we already determined
 and at fixed values of $s_{0}$.

 From these figures we obtained that the prediction of  the magnetic moment $\mu_{\Xi_{b}}$
 ($\mu_{\Xi_{c}}$) is practically independent of the value of the auxiliary parameter $\beta$. From
 all these analysis we deduce  the final results for the
 magnetic moments in Table 1 for $\chi=-3.15~GeV^2$.
Comparison of our results on the magnetic moments of $\Xi_{Q}$ baryons with predictions of other approaches, as we already noted, is also presented in
Table1.
\begin{table}[h]
\centering
\begin{tabular}{|c|c|c|c|c|} \hline
 &$\mu_{\Xi_{b}^{0}}$ & $\mu_{\Xi_{b}^{-}}$& $\mu_{\Xi_{c}^{0}}$&$\mu_{\Xi_{c}^{+}}$ \\\cline{1-5}
Our results &$-0.045\pm0.005$&$-0.08\pm0.02$ &$0.35\pm0.05 $&$0.50\pm0.05 $\\\cline{1-5}
RQM \cite{Faessler}&-0.06 &-0.06 &0.39&0.41\\\cline{1-5}
NQM \cite{Faessler}&-0.06&-0.06&0.37&0.37\\\cline{1-5}
\cite{Patel}&-&-&$-1.02\div-1.06$&$0.45\div0.48$\\\cline{1-5}
\cite{Savage}&-&-&0.32&0.42\\\cline{1-5}
\cite{DOR}&-&-&0.38&0.38\\\cline{1-5}
\cite{Oh}&-&-&0.28&0.28\\\cline{1-5}
\cite{An}&-&-&$0.28\div0.34$&$0.39\div0.46$\\\cline{1-5}
 \end{tabular}
 \vspace{0.8cm}
\caption{Results for the   magnetic moments of $\Xi_{Q}$ baryons in different approaches.
}\label{tab:1}
\end{table}

We see that within errors our predictions on the magnetic moments are in
good agreement with the quark model predictions. Our results on the magnetic moments of $\Xi_{c}$ are also close to the predictions of the other approaches  except the prediction of \cite{Patel} on $\mu_{\Xi_{c}^{0}}$.

In summary, the magnetic moments of $\Xi_{Q}$ baryons, which
were discovered recently (more precisely $\Xi_{b}$ was discovered) are
calculated in framework of light cone QCD sum rules. Our results on
magnetic moments are close to the predictions of the other approaches existing in the  literature.
\section{Acknowledgment}
Two of the authors (K. A. and A. O.), would like to thank TUBITAK,
Turkish Scientific and Research Council, for their partial financial
support both through the scholarship program and also through the
project number 106T333. One of the authors (A. O. ) would like to thank TUBA for funds provided through the GEBIP program.

\clearpage
 \begin{figure}[h!]
\begin{center}
\includegraphics[width=13cm]{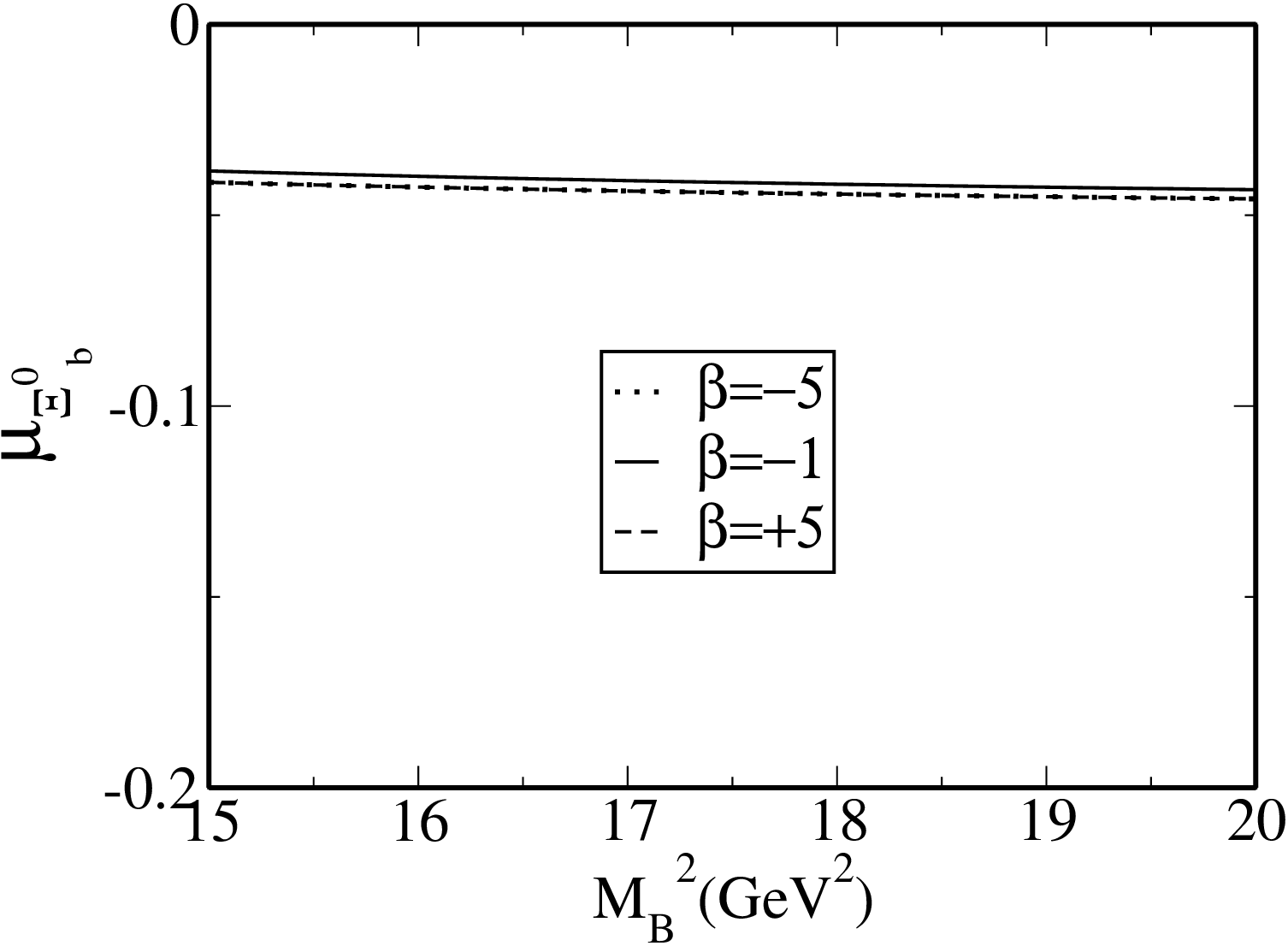}
\end{center}
\caption{The dependence of magnetic moment $\mu_{\Xi_{b}^{0}}$ on
$M_{B}^{2}$ at  $s_{0}=6.5^2~GeV^2$ and $\beta=\pm5,~-1$.} \label{fig1}
\end{figure}
\begin{figure}[h!]
\begin{center}
\includegraphics[width=13cm]{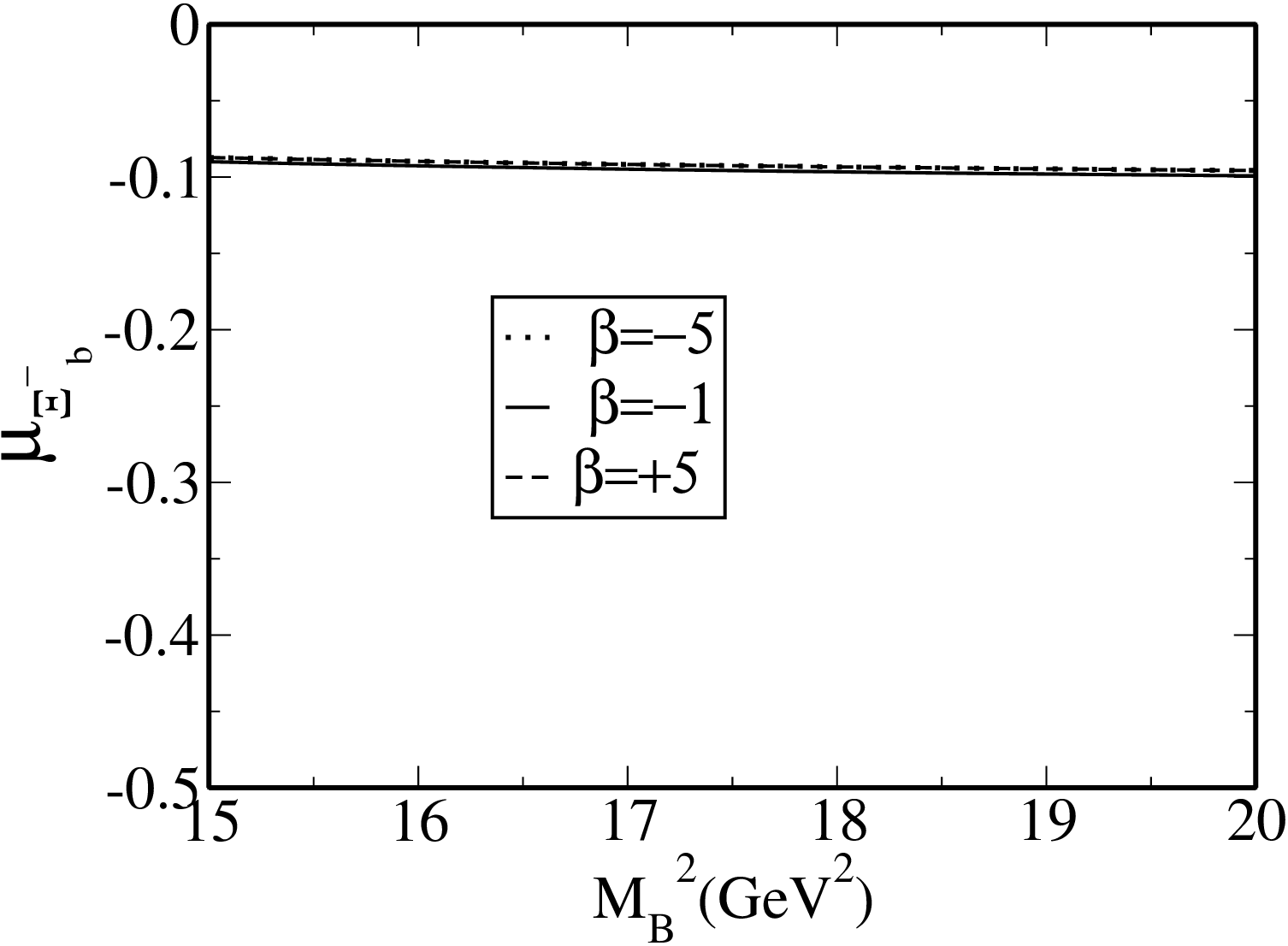}
\end{center}
\caption{The same as Fig. 1 but for $\mu_{\Xi_{b}^{-}}$.}
\label{fig2}
\end{figure}
\begin{figure}[h!]
\begin{center}
\includegraphics[width=13cm]{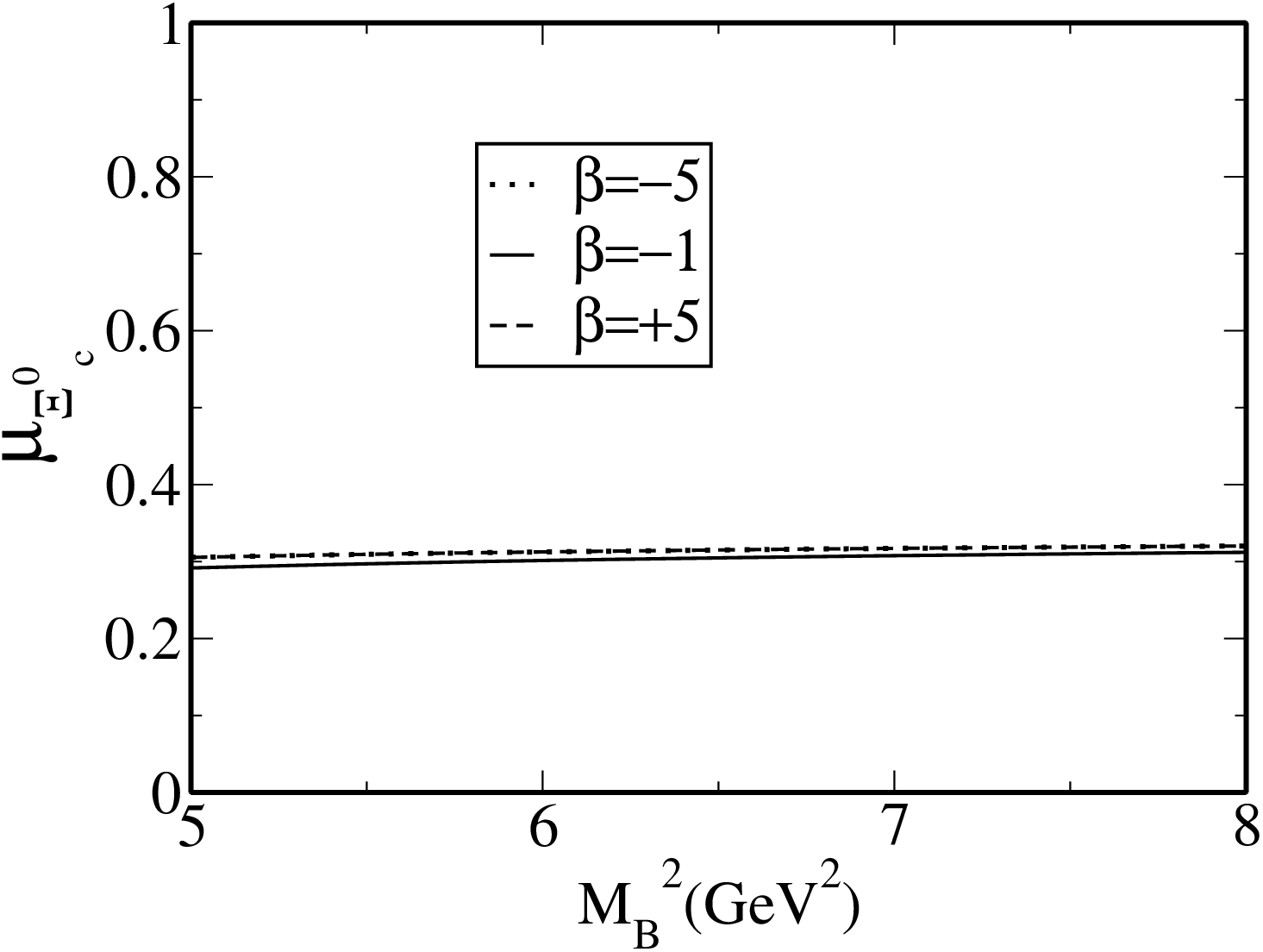}
\end{center}
\caption{The same as Fig. 1 but for $\mu_{\Xi_{c}^{0}}$ and at $s_{0}=3.0^2~ GeV^2$ .}
\label{fig3}
\end{figure}
\begin{figure}[h!]
\begin{center}
\includegraphics[width=13cm]{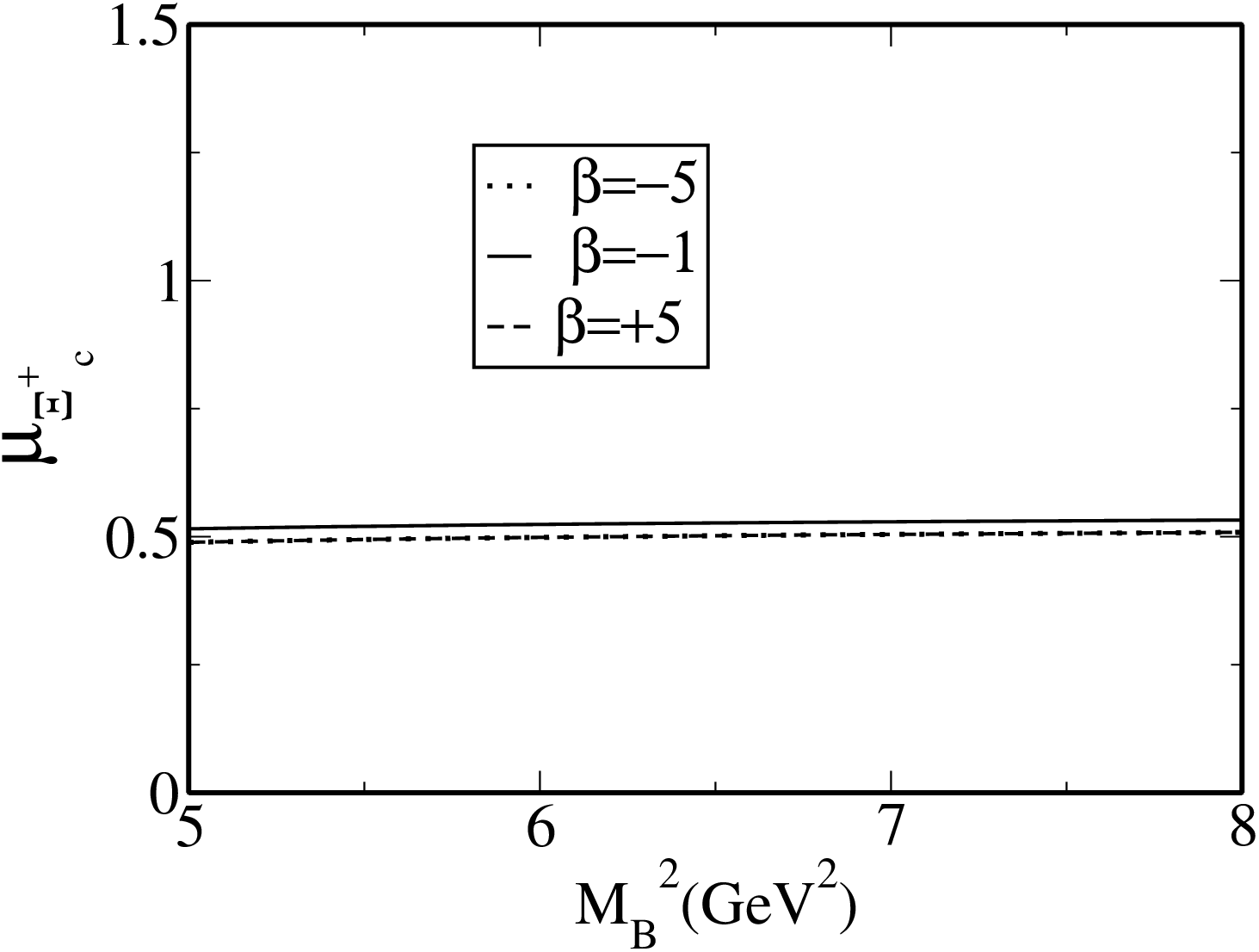}
\end{center}
\caption{The same as Fig.3 but for $\mu_{\Xi_{c}^{+}}$.}
\label{fig4}
\end{figure}
\begin{figure}[h!]
\begin{center}
\includegraphics[width=13cm]{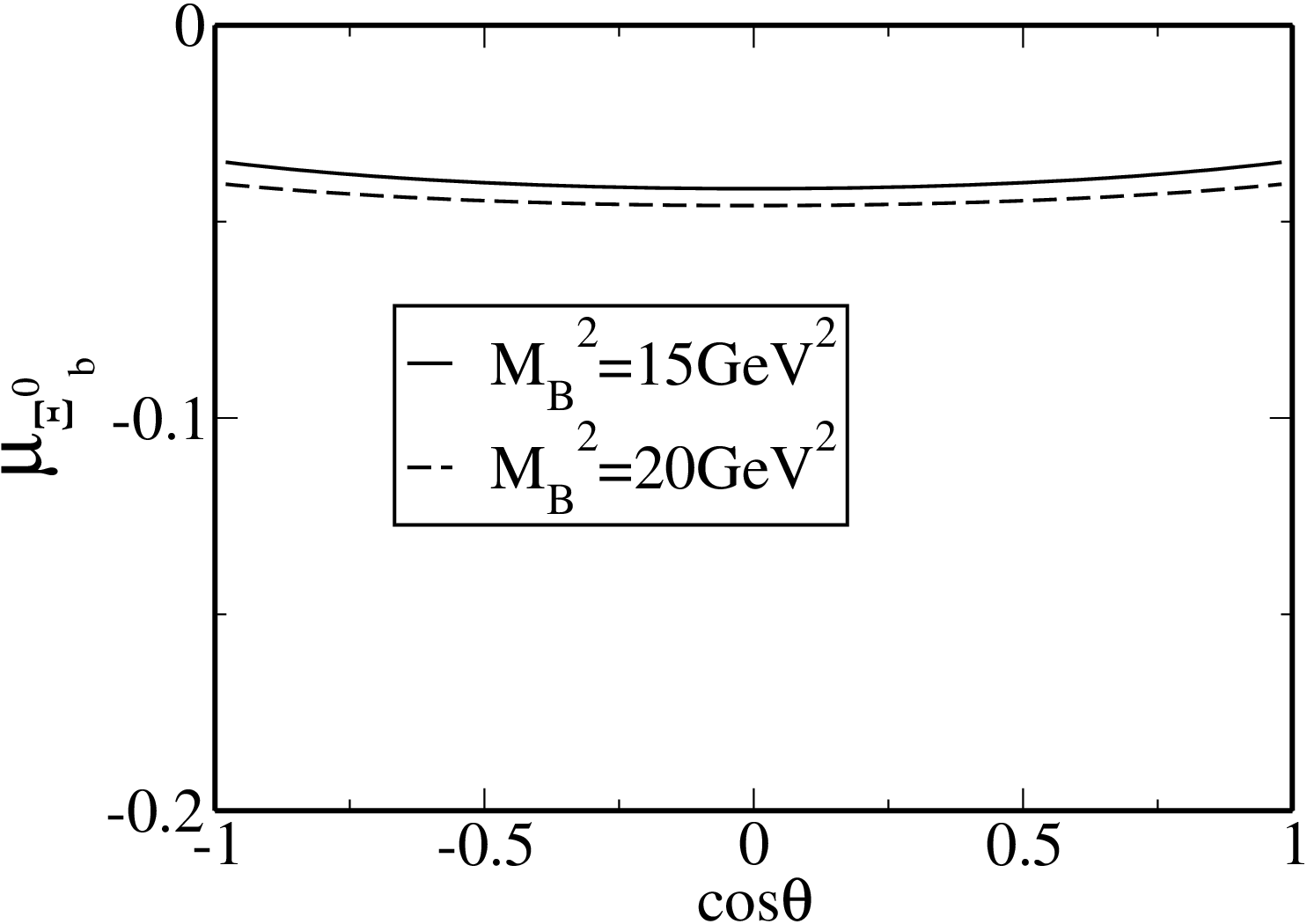}
\end{center}
\caption{The dependence of the  magnetic moment $\mu_{\Xi_{b}^{0}}$ on
$cos\theta$ at  $s_{0}=6.5^2~ GeV^2$ and for $M_{B}^2=15~GeV^2$ and $M_{B}^2=20~GeV^2$.}\label{fig5}
\end{figure}
\begin{figure}[h!]
\begin{center}
\includegraphics[width=13cm]{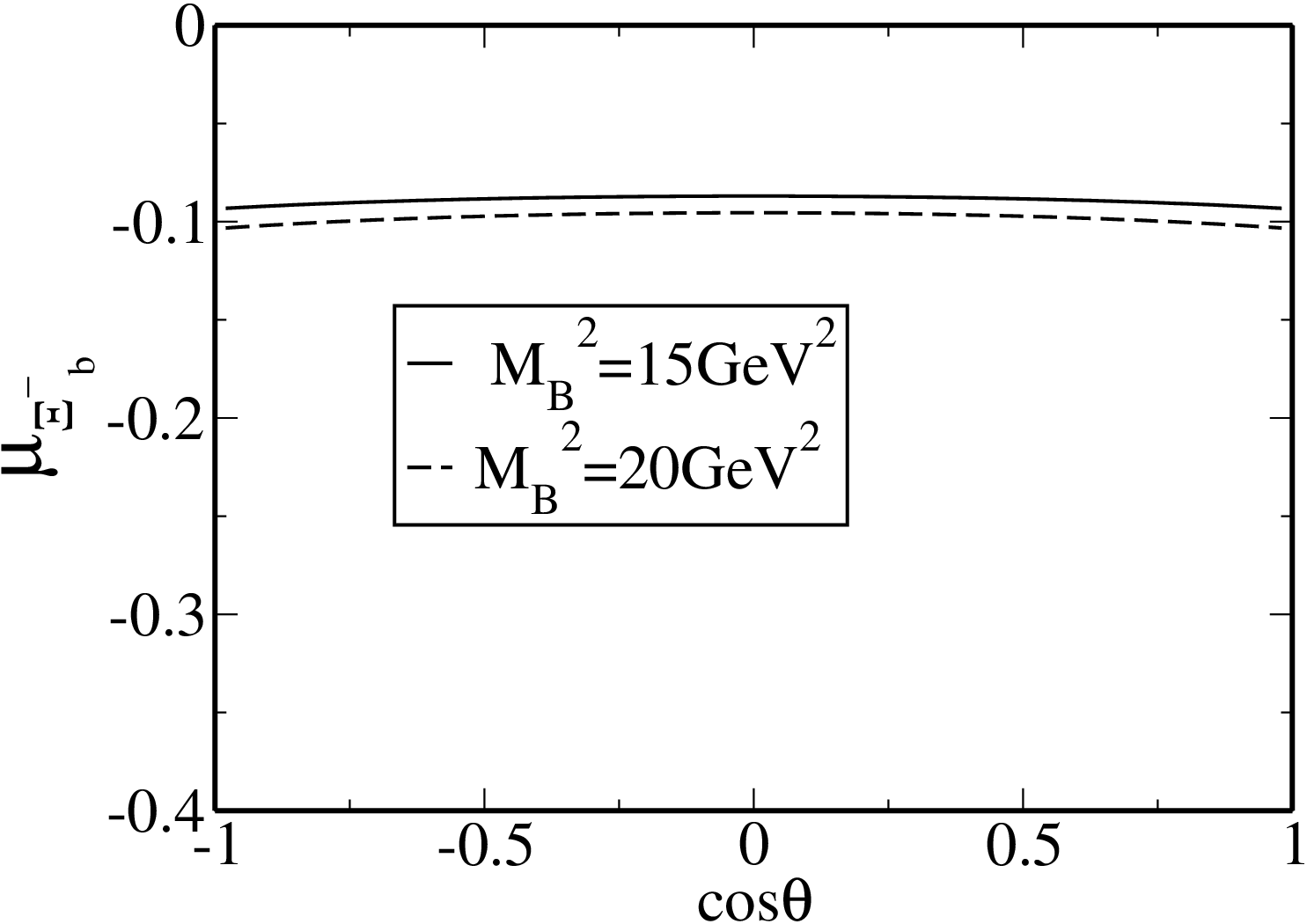}
\end{center}
\caption{The same as Fig. 5 but for $\mu_{\Xi_{b}^{-}}$.}
\label{fig6}
\end{figure}
\begin{figure}[h!]
\begin{center}
\includegraphics[width=13cm]{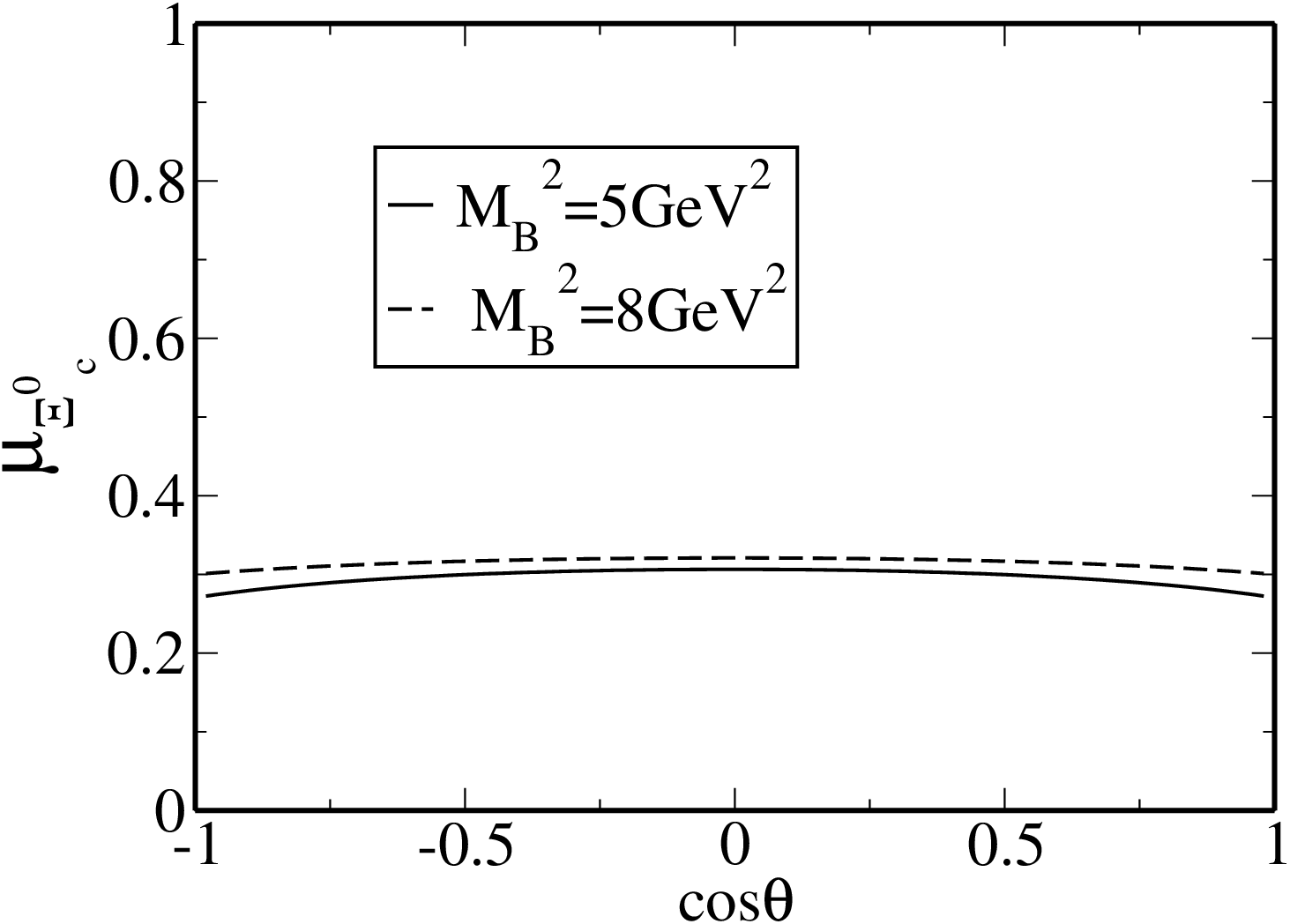}
\end{center}
\caption{The same as Fig. 5 but for $\mu_{\Xi_{c}^{0}}$ and $s_{0}=3.0^2~ GeV^2$ and for $M_{B}^2= 5~GeV^2$ and $M_{B}^2= 8~GeV^2$.}
\label{fig7}
\end{figure}
\begin{figure}[h!]
\begin{center}
\includegraphics[width=13cm]{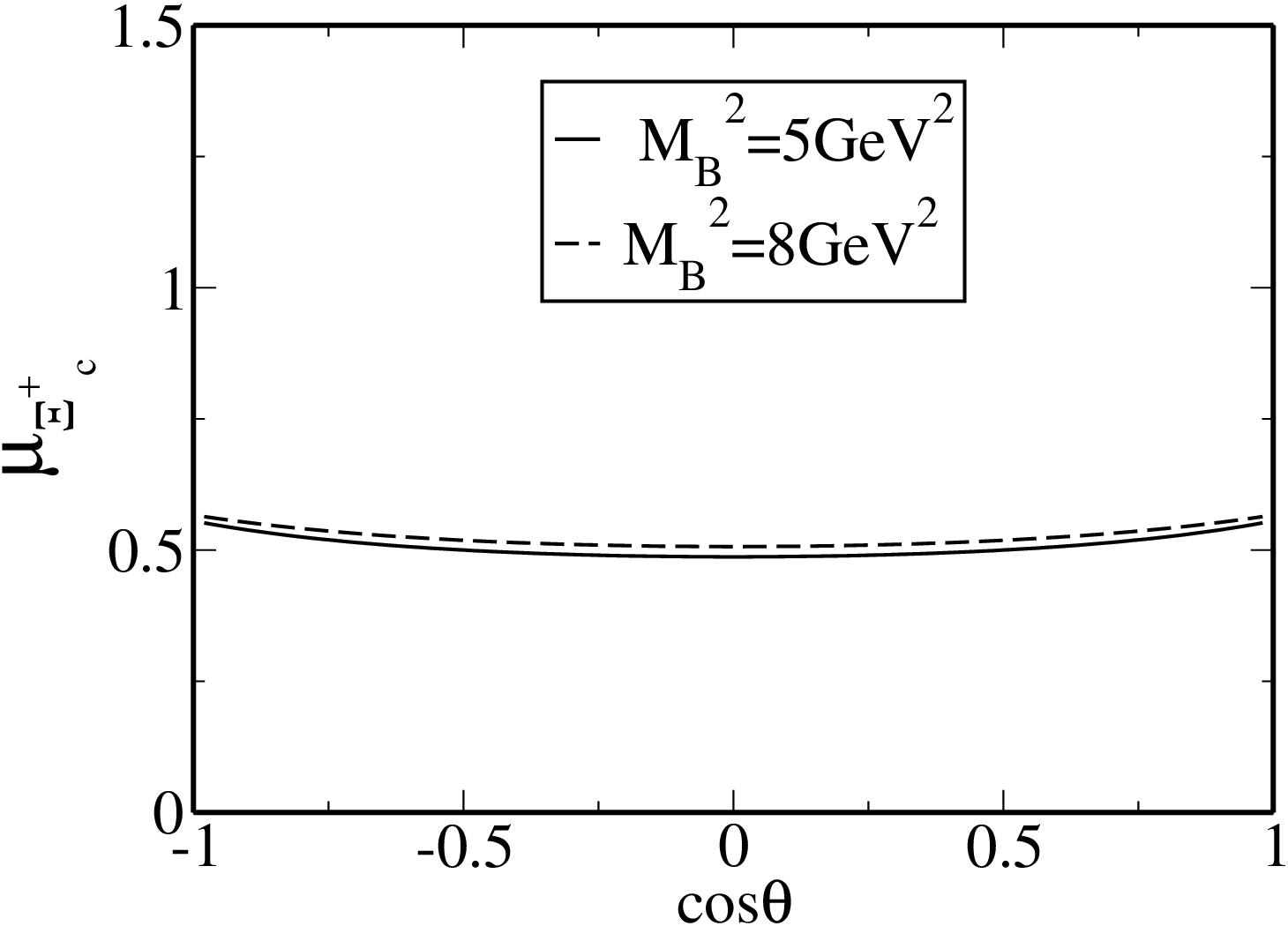}
\end{center}
\caption{The same as Fig. 7 but for $\mu_{\Xi_{c}^{+}}$.}
\label{fig8}
\end{figure}
\end{document}